\documentclass{aa}  

\usepackage{graphicx}	
\usepackage{amsmath}	
\usepackage{txfonts}
\usepackage{bm}
\newsavebox{\measurebox}
\usepackage{subcaption}
\usepackage{color}
\usepackage{ulem}
\usepackage{adjustbox}
\usepackage{tablefootnote}
\usepackage{threeparttable}
\usepackage[justification=raggedright]{caption}
\usepackage{booktabs}
\usepackage{tabularx}
\usepackage{soul}
\usepackage{gensymb}
\usepackage[dvipsnames]{xcolor}
\usepackage{multicol}
\usepackage{multirow}
\usepackage{placeins}
\newcommand{\Chandra}{\textit{Chandra}\xspace}

\newcommand{\Swift}{\textit{Swift}\xspace}

\usepackage{hyperref}

\begin{document}

   \title{A \Chandra view of SPT-CL J0217-5014: a massive galaxy cluster at a cosmic intersection at $z=0.53$}

   \author{Dan Hu \inst{1},
           Shida Fan \inst{2},
           Zhongsheng Yuan \inst{3},
           Junjie Mao \inst{4},
           Norbert Werner \inst{1},
           Yuanyuan Su \inst{5},
           François Mernier \inst{6}, 
           Yuanyuan Zhao \inst{2},
           Liyi Gu \inst{7}    
           \and
           Haiguang Xu \inst{2}
          }

   \institute{
        Department of Theoretical Physics and Astrophysics, Faculty of Science, Masaryk University, Kotl\'{a}\v{r}sk\'{a} 2, Brno 611 37, Czech Republic \\
        \email{hudan.bazhaoyu@mail.muni.cz}
        \and  
        School of Physics and Astronomy, Shanghai Jiao Tong University, Dongchuan Road 800, Shanghai 200240, China
        \and
        National Astronomical Observatories, Chinese Academy of Sciences, 20A Datun Road, Beĳing 100101, P. R. China
        \and
        Department of Astronomy, Tsinghua University, Beijing 100084, People's Republic of China \\
        \email{jmao@mail.tsinghua.edu.cn}
        \and
        Department of Physics and Astronomy, University of Kentucky, 505 Rose Street, Lexington, KY 40506, USA
        \and 
        NASA Goddard Space Flight Center, Code 662, Greenbelt, MD 20771, USA
        \and
        SRON Netherlands Institute for Space Research, Niels Bohrweg 4, 2333 CA Leiden, The Netherlands
        }

   \date{Received ; accepted }

 
  \abstract
   { Galaxy clusters trace the densest regions of the cosmic web and are crucial laboratories for studying the thermodynamic and chemical evolution of the intracluster medium (ICM). A massive galaxy cluster SPT-CL J0217-5014 ($z \sim 0.53$; $M_{\rm 500} \sim 3 \times 10^{14}~\rm M_{\sun}$) is one of the \Swift X-Ray Telescope serendipitous galaxy clusters with the highest reported Fe abundance ($\sim 1.3\pm 0.4$~$\rm Z_{\sun}$ within $\sim 1\arcmin.7$) and a potentially disturbed morphology. }
   { SPT-CL J0217-5014 presents an intriguing opportunity to investigate ICM chemical enrichment and cool-core survival. This study aims to evaluate its chemical and thermodynamic properties with a dedicated \Chandra observation.}
   { Using new \Chandra observations, we derived surface brightness profiles and dynamical state parameters. We also performed spectral fitting using different backgrounds to constrain the Fe abundance. Joint analysis of X-ray surface brightness, temperature, and integrated Sunyaev-Zel'dovich Compton parameter was performed to constrain the density profile. The DESI optical galaxy cluster catalogue was examined to explore its large-scale environment. }
   { The X-ray morphology reveals a disturbed ICM with a surface brightness edge at $\sim 0\arcmin.26$ ($\sim 100$~kpc) to the west and a tail-like feature extending towards the east. The best-fit metal abundance within $1\arcmin.5$ ($\sim 0.7\rm R_{500}$) is $0.61_{-0.23}^{+0.26}~\rm Z_{\sun}$. The derived central electron number density, entropy, and cooling time classify this system as a non-cool-core cluster, suggesting that merger activity has likely disrupted the possible pre-existing cool core. At larger radii ($\sim 1\arcmin - 2\arcmin$), we detect excess X-ray emission to the south, spatially aligned with a filamentary distribution of red galaxies, indicating ongoing accretion along an intracluster filament. Based on the DESI DR9 cross-matched optical clusters and photometric redshifts, we identify three nearby, lower-mass clusters that likely trace the large-scale structures, suggesting that SPT-CL~J0217-5014 is the primary node of a dynamically active environment where past mergers and anisotropic accretion along cosmic filaments have shaped the present-day ICM.}
  {}

   \keywords{ galaxies: clusters: general -- galaxies: clusters: individual: SPT-CL J0217-5014 -- X-rays: galaxies: clusters}

    \titlerunning{Galaxy cluster SPT-CL J0217-5014 at $z\sim 0.53$ }
    \authorrunning{Hu et al.}   

   \maketitle
%

\section{Introduction}
\label{sec:Sec1}

Galaxy clusters, the most massive gravitationally bound systems in the Universe, are thought to form at the intersections of filaments within the cosmic web. They grow through mergers and the accretion of surrounding lower-mass systems \citep[e.g.,][]{KB12}. These assembly processes, driven by the gravitational collapse and merger-induced energy release, heat the intracluster medium (ICM) to temperatures exceeding $10^7$~K, producing thermal bremsstrahlung radiation that dominates in the X-ray band \citep[e.g.,][]{sarazin86,BW10}. 

Observations and simulations suggest that proto-clusters or early-stage galaxy cluster assemblies were already in place at redshifts $z \sim 2-7$ \citep[e.g.,][]{daddi09,chiang13,miller18,esposito25}. Within these structures, galaxies experienced intense starburst activities, leading to the formation of massive stars that exploded as core-collapse supernovae (SNcc) on timescales of $\sim 10^6 - 10^7$~yr. These supernovae enriched the surrounding medium with $\alpha$-elements (e.g., O, Mg; \citealt{nomoto13}), which were subsequently dispersed outward by active galactic nucleus (AGN) feedback, supernova-driven winds, and stellar winds (see \citealt{biffi18}, for a review). Over time, this metal-enriched gas mixed into the ICM, resulting in a relatively flat and uniform metallicity distribution in cluster outskirts, typically around $0.3~\rm Z_{\sun}$ \citep[e.g.,][]{fujita08,werner13,simionescu15,urban17,mernier18,lovisari19}.

At lower redshifts ($z<1$), galaxy clusters are believed to approach virial equilibrium. During this phase, the central brightest cluster galaxy (BCG) could play an important role in enriching the ICM via stellar mass loss \citep{bohringer04,werner06,mao19} and supernova explosions (\citealt{werner08,mernier18}), with feedback from the AGN in the BCG helping to distribute iron enriched gas throughout the cluster core. Interestingly, most clusters exhibiting prominent central iron peaks are found to be cool-core systems \citep[e.g.,][]{grandi01,ettori15,mcdonald16,mantz17,lovisari19,liu20}, as their dense, low-entropy gas suppresses turbulent mixing and efficiently retains enriched material in the centre. Minor mergers may induce gentle gas sloshing that redistributes metals within the core region, but generally do not erase the central peak \citep[e.g.,][]{roediger11,ghizzardi14,hu19}. While major mergers or powerful AGN outbursts have the potential to disrupt the core and flatten the metallicity gradient, whether a major merger can fully erase the central metallicity peak remains debated. Some simulation studies suggest that the peak can survive or re-establish on $\sim$Gyr timescales, depending on the merger geometry and feedback history \citep[e.g.,][]{vogelsberger18,VS21}. 
Furthermore, some statistical analyses of cluster samples across a wide redshift range ($0<z<1.5$) suggest a mild increase in central iron abundance from $z\sim1$ to $z\sim0$ \citep[e.g.,][]{ettori15,mantz17,liu20}.
This trend supports the interpretation that the central abundance peak results from long-term, sustained enrichment from the BCG combined with the stable, dense environment of cool-core clusters. 
However, these samples are dominated by low-redshift systems ($z<0.3$), and robust measurements at $z>0.5$ remain limited.

A particularly notable example is WARP J1415.1$+$3612 at $z \sim 1$, which was found to host a well-developed cool core and an exceptionally high central abundance of $Z_{\rm Fe} = 3.6~\rm Z_{\odot}$ within $r<12$~kpc, along with evidence of star formation ($2-8~\rm M_{\sun}~yr^{-1}$) and AGN activity ($L_{\rm 1.4~GHz} = 2 \times 10^{25}$~$\rm W~Hz^{-1}$) in its BCG \citep{santos12}. This discovery indicated that the ICM had been rapidly enriched by early star formation in the BCG at $z > 1$ and feedback processes were already in place at $z \sim 1$. 
\cite{degrandi14} confirmed the presence of a central iron excess in the WARP J1415.1$+$3612 cluster, and further proposed that this iron excess resulted from early and intense star formation that happened around $z\sim3$, followed by SNIa explosions on a very short timescale ($\sim 0.04$~Gyr).

Similarly, as one of the \Swift X-Ray Telescope (XRT) serendipitous galaxy clusters, SPT-CL J0217-5014 was reported to exhibit a high gas temperature ($\sim 8.0\pm1.0$~keV) and super-solar iron abundance ($\sim 1.3 \pm 0.4~\rm Z_{\sun}$) within $1\arcmin.7$ ($\sim 0.8\rm R_{500}$; \citealt{tozzi14}). They also determined a redshift of $z \sim 0.52 \pm 0.01$ by fitting the Fe~K$\alpha$ line complex, which is consistent with other redshift estimates from optical observations \citep{bleem15,bocquet19,zou21,zou22}. 
The detection of high abundance over a large radial scale ($\sim 0.8\rm R_{500}$) suggests that the cluster SPT-CL J0217-5014 may be an outlier compared to the typical redshift-abundance trend derived from central ($0-0.1\rm R_{500}$) or intermediate ($0.1-0.5\rm R_{500}$) regions \citep{ettori15,mantz17,liu20}. In addition, the dynamical state and thermodynamic distribution of this galaxy cluster remain poorly understood.
To confirm whether this serendipitously discovered cluster indeed has a high abundance, potentially providing us with another "J1415.1$+$3612"-like target, and to deepen our understanding of this source, we obtained a first \Chandra observation with about 100~ks exposure time (PI: J. Mao). The cluster’s mass, derived from Sunyaev–Zel'dovich (SZ) effect measurements, is estimated to be $M_{500} \sim 3 \times 10^{14}~M_\odot$ \citep{bleem15,bocquet19}. According to the photometric redshift ($z=0.53 \pm 0.01$) of the BCG of SPT-CL J0217-5014 from DESI DR9 data \citep{zou22}, $R_{500}$, where the enclosed matter density is 500 times the critical density at this redshift, is estimated as $850$~kpc by adopting $M_{500}=\frac{4\pi}{3}R_{500}^{3}500\rho_{\rm c}(z)$, where $\rho_{\rm c}(z) = 3 H(z)^2/8\pi G$ is the critical density. 

Throughout this work, we adopt a standard flat $\Lambda$CDM cosmology with $\Omega_{m}=0.27$, $\Omega_{\lambda}=0.73$, and the Hubble constant $H_{0}=70~\rm km~s^{-1}~Mpc^{-1}$. At the redshift of SPT-CL J0217-5014 ($z=0.53$), the angular scale of the target is $\sim 6.36~\rm kpc/\rm arcsec$, and the angular diameter distance and luminosity distance are 1312~Mpc and 3072~Mpc, respectively. Unless otherwise stated, we adopt the solar abundance standards of \cite{wilms00} and quote errors at the 68\% confidence level.

\section{\Chandra observations and data reduction}
\label{sec:sect2}

SPT-CL J0217-5014 was observed with the Advanced CCD Imaging Spectrometer (ACIS)-I array on board \Chandra during three pointing observations, performed on March 11, 2020 (ObsID: 21538; 45.5~ks), March 17, 2020 (ObsID: 23045; 29.7~ks), and March 18, 2020 (ObsID: 23191; 30.7~ks).  
For each observation, the data was reduced following the standard procedure as suggested by the \Chandra X-ray Center, using the \Chandra Interactive Analysis of Observations (\textsc{ciao}; \citealt{fruscione06}) v4.16.0 and \Chandra Calibration Database (\textsc{caldb}) v4.10.2. The detailed data reduction procedure is provided in \citet{hu19,hu21}, we briefly introduce several main steps here. First, we used the \textsc{ciao} script \texttt{chandra\_repro} to reprocess level 1 event files by applying the latest calibration files and producing new bad pixel maps. Then, we applied the \textsc{ciao} tool \texttt{wavdetect} with \texttt{scale} parameters of 1.0, 2.0, 4.0, 8.0, 16.0, 24.0 and the significance threshold (\texttt{sigthresh}) of $10^{-6}$ to identify and exclude all the point sources detected in the ACIS images for each observation. All the detected point sources were confirmed by visual inspection. We then examined the light curves extracted in the $0.5-12.0$~keV band from the source-free region. Time intervals contaminated by occasional particle background flares during which the count rate deviates from the mean value by 20\% were excluded. The final total clean exposure time of these three observations is 104.5~ks. 

Three types of background files, including blank-sky background, stowed background, and local background, were used in this work.
The blank-sky background represents composite observations of relatively empty sky fields with point sources removed, combining both X-ray sky background and instrumental background. 
While the stowed background represents pure particle-induced instrumental background obtained when the ACIS detector is positioned away from the telescope's focal plane and shielded from all sky X-rays.
The background files were obtained using the \textsc{ciao} script \texttt{blanksky}. 
Two sets of local background regions were selected for each observation: one from an annular region ($3\arcmin - 4\arcmin$) and the other from three box regions away from the source, shown in Figure~\ref{fig:expcorr-img}(a). 

In this work, the spectra with the appropriate ancillary response files (\texttt{ARFs}) and redistribution matrix files (\texttt{RMFs}) were extracted using the \textsc{ciao} tool \texttt{specextract} and fitted by using \textsc{xspec} v12.13.0 \citep{arnaud96} and \textsc{atomdb} v3.0.9 \citep{smith01,foster12}.

\section{X-ray imaging analysis}
\label{sec:xray-img}

\subsection{X-ray surface brightness distributions}
\label{sec:sbp} 
%

\begin{figure*}
   \centering
   \includegraphics[scale=.35]{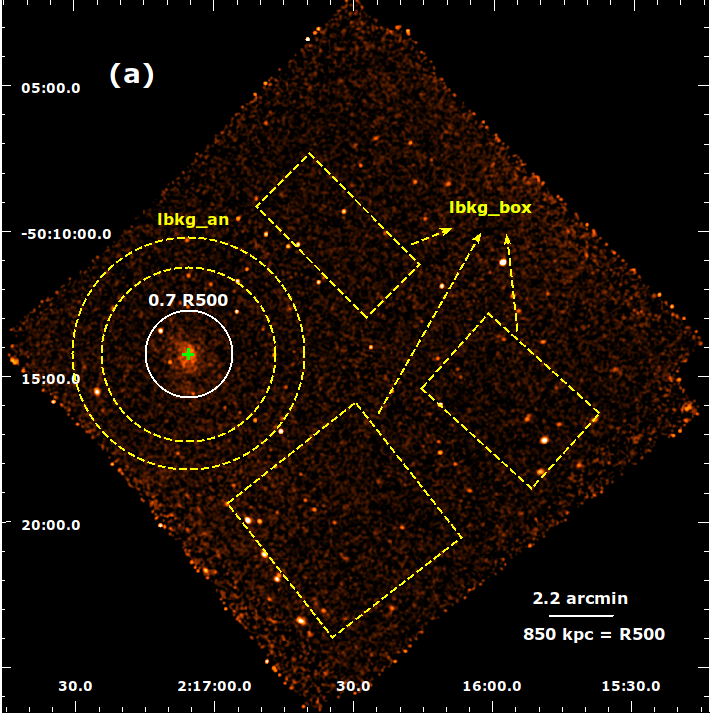}
    \includegraphics[scale=.35]{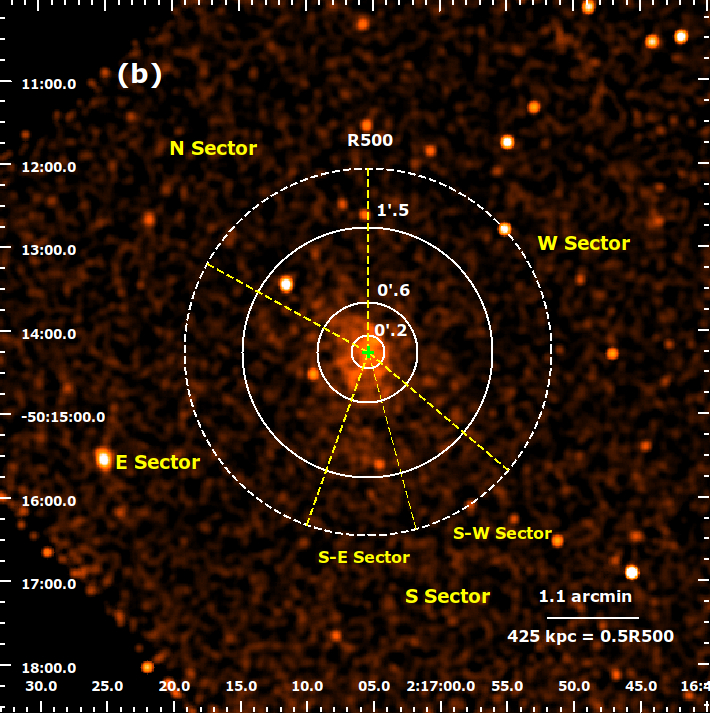}
    \caption{Exposure-corrected 0.5–7 keV Chandra ACIS-I0–3 image of SPT-CL J0217-5014. (a) The source extraction region, centred on the X-ray centroid, is shown by a white circle with a radius of $1'.5$ ($\sim 0.7R_{500}$). The regions used for local background extraction are also indicated. All point sources were excluded from both the source and background regions during imaging and spectral analysis. (b) Zoomed-in view of the cluster core. The annuli used for radial analysis, with radii of $0'.2$, $0'.6$, and $1'.5$, are shown, together with the four sectors used to extract surface brightness profiles. The S sector is further divided into two equal sectors.
    }
    \label{fig:expcorr-img}
\end{figure*}
%

The exposure-corrected $0.5-7$~keV ACIS I0-3 image of SPT-CL J0217-5014 generated using data from the three \Chandra observations and the \textsc{ciao} tool \texttt{merge\_obs} is shown in Figure~\ref{fig:expcorr-img}. The image was smoothed with a Gaussian kernel of 5 pixels. 
The X-ray centroid was estimated within a circular region of radius $1\arcmin.5$ and is located at $\rm R.A.=02^h 17^m 05.39^s, decl. =-50^\circ 14^\prime 16.2^{\prime \prime}$ (J2000.0), as marked with a cross in Figure~\ref{fig:expcorr-img}.
The position of the BCG is $\rm R.A.=02^h 17^m 05.50^s, decl. =-50^\circ 14^\prime 16.44^{\prime \prime}$ (J2000.0), with an offset of only $\sim 1\arcsec$ from the X-ray centroid.
We note that no diffuse radio emission has been detected in SPT-CL J0217-5014 with current radio surveys, possibly due to their limited sensitivity and poor integrated observation time. However, the central AGN associated with the BCG is detected in the Rapid ASKAP Continuum Survey (RACS) at 1367.5~MHz, with a radio luminosity of $L_{1.37~\rm GHz} = 1.3 \times 10^{24}$~$\rm W~Hz^{-1}$ \citep{duchesne24}.

\begin{figure*}
    \centering
    \includegraphics[width=0.95\textwidth]{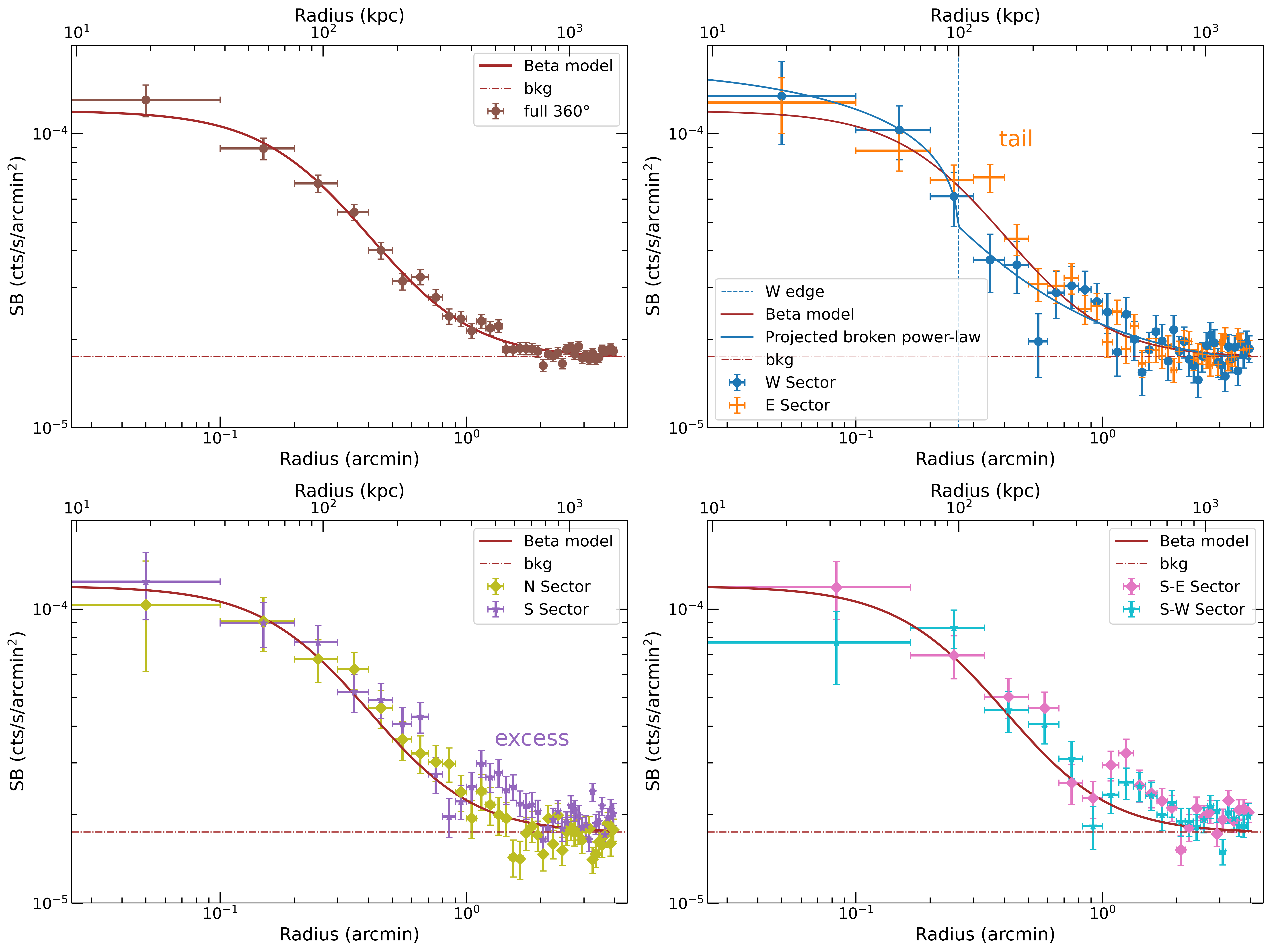}
    \caption{Radial surface brightness profiles extracted from a set of azimuthally averaged (full $360^\circ$) regions and from sector regions in four different directions, as indicated in the legend of each panel. The best-fit single-$\beta$ model for the azimuthally averaged surface brightness profile and the best-fit background parameter are also presented in all panels. The best-fit projected broken power-law model for the western surface brightness profile is presented in the top-right panel.}
    \label{fig:sbp}
\end{figure*}

Within the inner region ($< 0\arcmin.6$; see Figure~\ref{fig:expcorr-img}(b)), the X-ray emission shows an asymmetric morphology with several potential substructures, e.g., an arc-shaped edge west of the cluster centre. To obtain a basic understanding of the X-ray surface brightness distribution of SPT-CL J0217-5014, we extracted surface brightness profiles from a set of azimuthally averaged (full $360^\circ$) regions and from sector regions in four directions (Figure~\ref{fig:expcorr-img}). The resulting surface brightness profiles are shown in Figure~\ref{fig:sbp}. Overall, the surface brightness decreases outward and approaches the background level beyond $1\arcmin.5$ ($\sim 0.7\rm R_{500}$). 
We used a single-$\beta$ model to fit the azimuthally averaged surface brightness profile using \textsc{pyproffit} \citep{eckert20} with a form of
\begin{equation}
    S(r) = S_0 \left(1 + (r/r_{\rm c})^2\right) ^ {-3 \beta + 0.5} + S_{\rm bkg}
\end{equation}
where $r_{\rm c} = 0\arcmin.34 \pm 0\arcmin.06$ is the core radius, $\beta = 0.58 \pm 0.06$ is the slope,  $S_0 = (7.7 \pm 1.0) \times 10^{-5}$~$\rm cts~s^{-1}~arcmin^{-2}~cm^{-2}$ is the normaliztion, and $S_{\rm bkg} = (1.72 \pm 0.02) \times 10^{-5}$~$\rm cts~s^{-1}~arcmin^{-2}~cm^{-2}$ is the background. The best-fit result is shown in Figure~\ref{fig:sbp} and indicates that there is no obvious excess at the cluster core.   

Along the western direction, an edge (W edge) is obvious at $\sim 0\arcmin.2$ ($\sim 77$~kpc), possibly corresponding to the cold front (or cool core). To characterise the W edge, the W surface brightness profile was modelled by a projected three-dimensional density profile implemented in \textsc{pyproffit} with a broken power-law form:
\begin{equation}
    n_{\rm e}(r) =
        \begin{cases}
            Cr^{-\alpha_{1}}  &  \text{if } r \leq r_{\rm break} \\
            C\frac{1}{d_{\rm jump}}r^{-\alpha_{2}} &  \text{if } r > r_{\rm break} ,
        \end{cases}
\end{equation}
Where $\alpha_{1}$ and $\alpha_{2}$ are the inner and outer slope indices at the break radius $r_{\rm break}$, and $d_{\rm jump} = n_{\rm e, in}/n_{\rm e, out}$ represents the density jump between the inner ($n_{\rm e, in}$) and outer ($n_{\rm e, out}$) regions. 
The best fit indicates a density jump of $d_{\rm jump} = 1.7 \pm 0.7$ at $r_{\rm break} = 0\arcmin.26 \pm 0\arcmin.03$.
In the eastern direction, a surface brightness discontinuity is also evident at $\sim 0\arcmin.4$ ($\sim 154$~kpc), which may represent an X-ray tail stripped from the cluster core. 
On larger scales, we detect an excess of X-ray emission between $\sim 1\arcmin - 2\arcmin$ to the south. 
This southern excess is spatially aligned with the filamentary distribution of red galaxies, potentially suggesting ongoing accretion along a south–north intracluster filament (see the optical image in Figure~\ref{fig:optical} and the discussion in Section~\ref{sec:DESI}).

\subsection{Dynamical state }
\label{sec:dyn} 
%

\begin{figure}
    \centering
    \includegraphics[scale=.6]{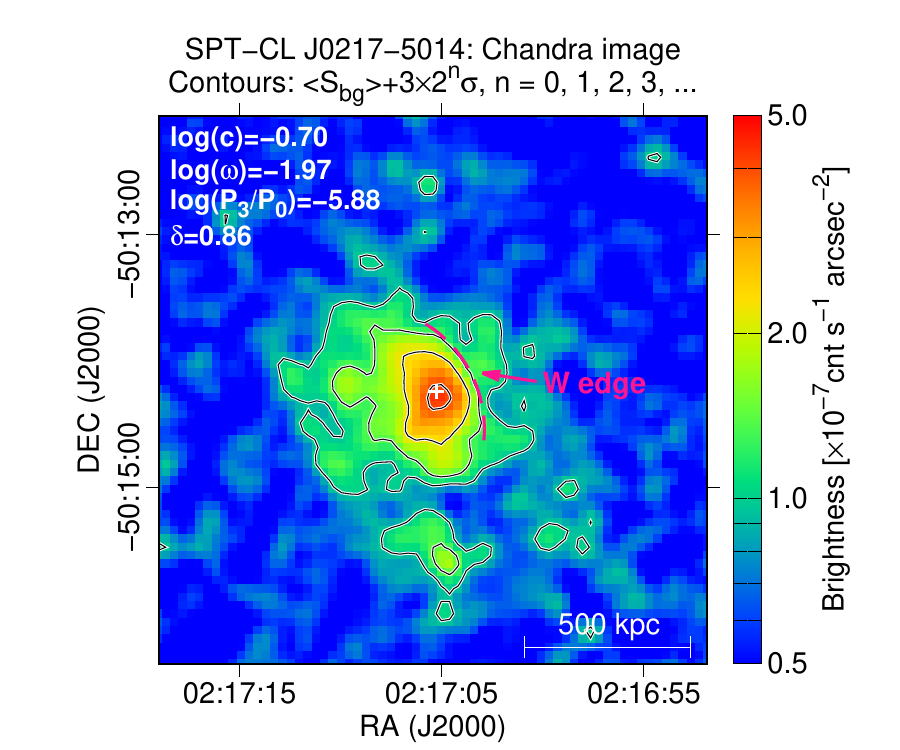}
    \caption{Brightness map of SPT-CL J0217-5014 with contours, the best-fit parameters of the dynamical state are shown in the top left corner. The W edge was also marked.}
    \label{fig:dyn}
\end{figure}

\begin{figure*}
    \centering
    \includegraphics[width=0.95\textwidth]{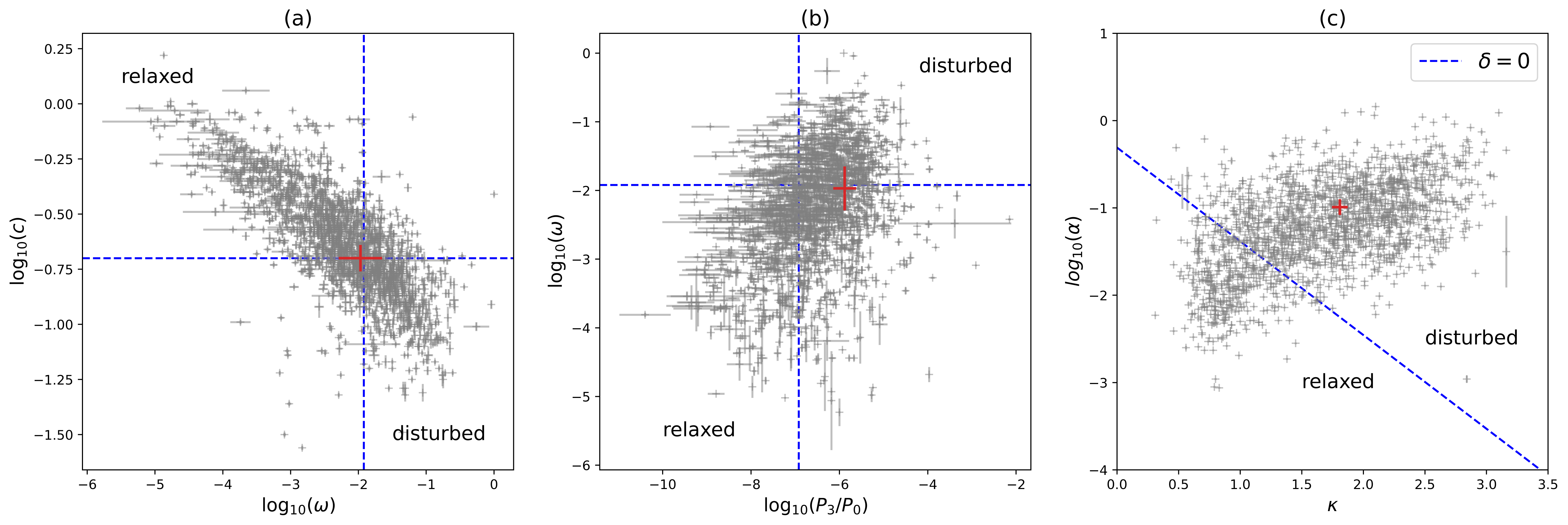}
    \caption{Distribution of 1844 galaxy clusters in three morphological parameter spaces (from left to right): $\omega - c$, $P_3/P_0 - \omega$, and $\kappa - \alpha$ (data were taken from \citealt{YHW22}). The position of SPT-CL J0217-5014 is added and highlighted in a red marker.}
    \label{fig:dyn-sample}
\end{figure*}

The asymmetric morphology shown in the exposure-corrected \Chandra image, along with the edges and excess indicated from the surface brightness profiles, suggests that the cluster SPT-CL J0217-5014 is dynamically disturbed. To further assess and quantify its dynamical state, we measured three widely used morphological parameters and a recently proposed parameter introduced by \cite{YH20}, following their method.  
\begin{enumerate}
    \item Concentration index ($c$):
Relaxed galaxy clusters typically host more luminous cool cores than disturbed galaxy clusters \citep[e.g.,][]{fabian84,mcdonald12}. Based on this, the concentration index $c$, defined as the ratio of surface brightness within a core radius of 100~kpc to that within a large radius of 500~kpc, can be used as an indicator of the dynamical state of a galaxy cluster. Following the method adopted in \cite{santos08} and \cite{cassano10}, the concentration index $c$ is calculated as 
\begin{equation}
    c = \frac{S(r<100~{\rm kpc})}{S(r<500~{\rm kpc})}
\end{equation}
This parameter is widely used to identify cool-core clusters, particularly at high redshifts, where spatially resolved spectroscopic analysis of the cluster core is often not feasible.
    \item Centroid shift ($\omega$): 
In merging galaxy clusters, the X-ray surface brightness peak often deviates significantly from the model-fitted centroid, in contrast to relaxed clusters, where the offset is typically small. \cite{poole06} quantified this offset using the centroid shift ($\omega$), defined as the standard deviation of the projected separation between the X-ray peak and the centroid within a series of concentric apertures ranging from $0.05R_{\rm ap}$ to $R_{\rm ap}$:
\begin{equation}
    \omega = \left[\frac{1}{n-1}\sum_{i}(\Delta_{i} - <\Delta>)^{2}\right]^{1/2} \times \frac{1}{R_{\rm ap}}
\end{equation}
where $n$ is the number of apertures, $\Delta_{i}$ is the distance between the X-ray peak and centroid in the $i$-th aperture, $<\Delta>$ is the mean of all $\Delta_{i}$, and $R_{\rm ap}$ is the maximum aperture radius. In our analysis, we adopted $R_{\rm ap} = 500$~kpc and $n=20$.    
    \item Power ratio (${P_3/P_0}$): 
The power ratio ${P_3/P_0}$ is derived from a multipole decomposition of the surface brightness map within a circular aperture and is particularly sensitive to asymmetric fluctuations and subclumps, which are more prominent in disturbed clusters \citep{BT95}. We refer to \cite{BT95} for the complete definition and formulation.
    \item Morphology index $\delta$: 
Given the fact that the above three traditional methods are prone to being affected by the distance (or redshift) of the galaxy clusters, \cite{YH20} introduced a new parameter, the morphology index $\delta$, as a redshift-independent indicator of cluster dynamic state. This index consists of two adaptive quantities derived from the \Chandra image: the profile parameter $\kappa=(1+\epsilon)/ \beta$, based on the ellipticity ($\epsilon$) and power-law index ($\beta$) of a fitted two-dimensional elliptical $\beta$-model, and the asymmetry factor $\alpha$, which quantifies the  deviations from point symmetry about the cluster centre:  
\begin{equation}
    \alpha = \frac{\sum\limits_{x_i, y_i}\left[f_{\rm obs}(x_i, y_i)-f_{\rm obs}(x_i^{'}, y_i^{'})\right]^2}{\sum\limits_{x_i,y_i}f_{\rm obs}^2(x_i, y_i)} \times 100~\rm per cent
\end{equation}
where $f_{\rm obs}(x_i, y_i)$ is the observed flux at pixel $(x_i,y_i)$, and $f_{\rm obs}(x_i^{'}, y_i^{'})$ is the flux at its point-symmetric counterpart with respect to the cluster centre $(x_0,y_0)$.
To construct the morphology index, \cite{YH20} fitted a linear equation of the form $ \delta = A\kappa + B\alpha + C$ to a sample of 125 galaxy clusters with known dynamical states. The best-fit empirical relation is:
\begin{equation}
    \delta = 0.68\rm log_{10}(\alpha) + 0.73\kappa + 0.21
\end{equation}
Clusters with $\delta >0$ are considered dynamically disturbed. This index correlates well with other morphological parameters while offering improved robustness against redshift \citep{YH20,YHW22}.
\end{enumerate}

By applying the four methods, we obtain the following measurements for SPT-CL J0217-5014: the concentration index $\mathrm{log_{10}}(c)=-0.70 \pm 0.06$, centroid shift $\mathrm{log_{10}(\omega)}= -1.97 \pm 0.32$, power ratio $\mathrm{log_{10}}(P_3/P_0) = -5.88 \pm 0.26$, and morphology index $\delta = 0.86 \pm 0.01$. The X-ray surface brightness map of SPT-CL J0217-5014 and derived morphological parameters are shown in Figure~\ref{fig:dyn}.
The classification criteria for the dynamical state of galaxy clusters are typically based on empirical thresholds derived from large samples. Following the empirical thresholds from \cite{cassano10} and \cite{YH20}, clusters with $\mathrm{log_{10}}(c) > -0.7$, $\mathrm{log_{10}(\omega)}<-1.92$, $\mathrm{log_{10}}(P_3/P_0) < -6.92$, and $\delta<0$, are generally considered relaxed or cool-core clusters.  
In figure~\ref{fig:dyn-sample}, the locus of SPT-CL J0217-5014 is marked in different morphological parameter spaces. We find that the power ratio $(P_3/P_0)$ and morphology index ($\delta$) clearly place SPT-CL J0217-5014 in the dynamically disturbed regime, likely because these parameters are more sensitive to substructures. 
In contrast, the concentration index ($c$) and centroid shift ($\omega$) lie close to the dividing thresholds, and thus do not provide a definitive classification regarding the presence of a cool core. 
Numerical simulations suggest that cool cores can, in some cases, survive minor or off-axis mergers \citep{poole08}, so the possibility of a residual weak cool core cannot be excluded.
Overall, the morphological indicators support a scenario in which SPT-CL J0217-5014 has undergone recent dynamical activity.

\section{X-ray spectral analysis}
\label{sec:spc} 

\subsection{Global spectral properties}
\label{sec:ave-spc} 

\begin{table}
    \caption{Spectral fitting results from different regions using the blank-sky background }
    \label{tab:spc1}
    \centering
    \renewcommand{\arraystretch}{1.4}
    \begin{threeparttable}
        \begin{tabular}{ccccc}
        \hline
        Radius  & Temperature & Abundance & C-stat/dof\\
        (arcmin) & (keV) & ($\rm Z_{\sun}$) & \\
        \hline   
        $0.0-1.5$ & $5.76_{-0.70}^{+0.78}$ & $0.61_{-0.23}^{+0.26}  $ &  1079/1208 \\  
        \hline      
        $0.0-0.2$  &  $6.57_{-1.62}^{+2.13}$ & $0.61$ (fixed) & 167/218 \\
        $0.2-0.6$  &  $9.66_{-2.06}^{+2.96}$ & $0.61$ (fixed)  & 462/597 \\
        $0.6-1.5$  &  $4.92_{-0.91}^{+1.17}$ & $0.61$ (fixed) & 978/1121 \\
        \hline
        \end{tabular}
    \end{threeparttable}
\end{table}
\begin{figure}
    \centering
    \includegraphics[scale=.22]{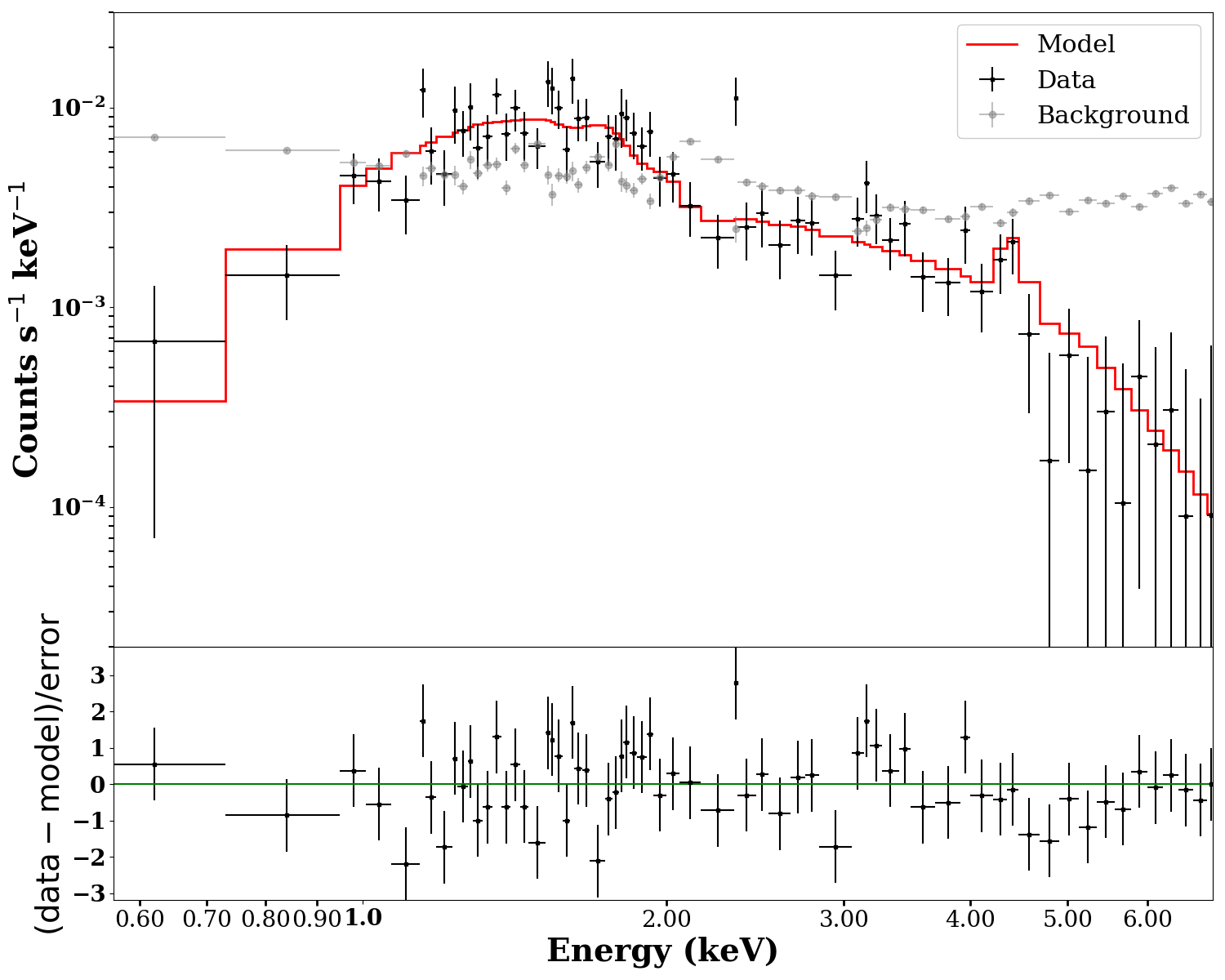}
    \caption{Best-fit X-ray spectra extracted within a $1\arcmin.5$ radius using the blank-sky background. The combined \Chandra spectrum has been binned to a minimum of 15 counts per bin, and is presented for illustrative purposes only. }
    \label{fig:spc_fit}
\end{figure}

As indicated by the azimuthally averaged surface brightness profile (top-left panel of Figure~\ref{fig:sbp}), the cluster surface brightness approaches the background level at around $1\arcmin.5$ ($\sim 0.7\mathrm{R}_{500}$). We therefore extracted spectra from a circular region with a radius of $1\arcmin.5$, centred on the X-ray centroid in each of the three observations separately, and analysed them by jointly fitting the spectra.
The background-subtracted spectra were fitted with an absorbed thermal model (\texttt{phabs*apec}), with the hydrogen column density $N_{\rm H}$ fixed at $1.63\times10^{20}~\rm{cm}^{-2}$ \citep{hi4pi16}. 
The gas temperature, metal abundance, and normalisation were left as free parameters. A grouping of one count per bin was applied, and the \texttt{C-statistic} \citep{kaastra17} was used for spectral fitting due to the limited photon statistics.  

The best-fit parameters obtained using the four different background files are consistent with each other. Here, we show only the results derived with the blank-sky background in Table~\ref{tab:spc1} and the best-fit spectrum in Figure~\ref{fig:spc_fit}. The corresponding results using three other background files are provided in Appendix~\ref{app-1}. The gas temperature is found to lie in the range $5-7$~keV. To reduce fitting uncertainties, the redshift was fixed at 0.53 during spectral analysis. We also tested leaving the redshift as a free parameter, yielding values in the range $z \sim 0.51 - 0.55$, consistent with earlier redshift measurements \citep{tozzi14,bleem15,bocquet19,zou22}. Although the derived metal abundance ($Z = 0.61_{-0.23}^{+0.26}$) is subject to large uncertainties, it remains lower than the supersolar value and is consistent with the typical redshift-abundance trend derived from central ($0-0.1~\rm R_{500}$) or intermediate ($0.1-0.5~\rm R_{500}$) regions \citep{ettori15,mantz17,liu20}.

We noticed that the gas temperature and abundance values derived within $1\arcmin.5$ ($\sim 0.7\rm R_{500}$) in this work are lower than those reported by \cite{tozzi14}, who used \Swift data and obtained a gas temperature of $8.03_{-0.93}^{+0.96}$~keV and a metal abundance of $1.26_{-0.34}^{+0.40}~\rm Z_{\sun}$ within $1\arcmin.73$ ($\sim 0.8\rm R_{500}$). 
To investigate this difference, we re-fitted the \Chandra spectrum using the blank-sky background, adopting the same abundance table (\texttt{Aspl}; \citealt{asplund09}) and models (\texttt{tbabs*mekal}) used in their analysis. We found that the gas temperature remained unchanged, and the abundance slightly changed to $0.49_{-0.19}^{+0.22}~\rm Z_{\sun}$.
This discrepancy could be attributed to the potential cross-calibration issues between \Chandra and \Swift XRT \citep[e.g.,][]{tsujimoto11,moretti12} and the possibility of residual point-source contamination in \Swift XRT data given its $\sim 15 \arcsec$ resolution. Moreover, the relatively higher non-X-ray background of \Chandra (compared to \Swift XRT; see \citealt{moretti07,moretti12}), may result in large residual uncertainties after background subtraction, reducing the significance of the Fe K lines relative to the continuum spectrum and affecting the measurements of both the bremsstrahlung component and the iron abundance.
Although the $1\sigma$ intervals of the measurements from \Chandra and \Swift do not overlap, they may still be marginally consistent at the 90\% confidence level.

\subsection{Radial spectral analysis and cool core identification}
\label{sec:cc} 

\begin{figure*}
    \centering
    \includegraphics[width=\textwidth]{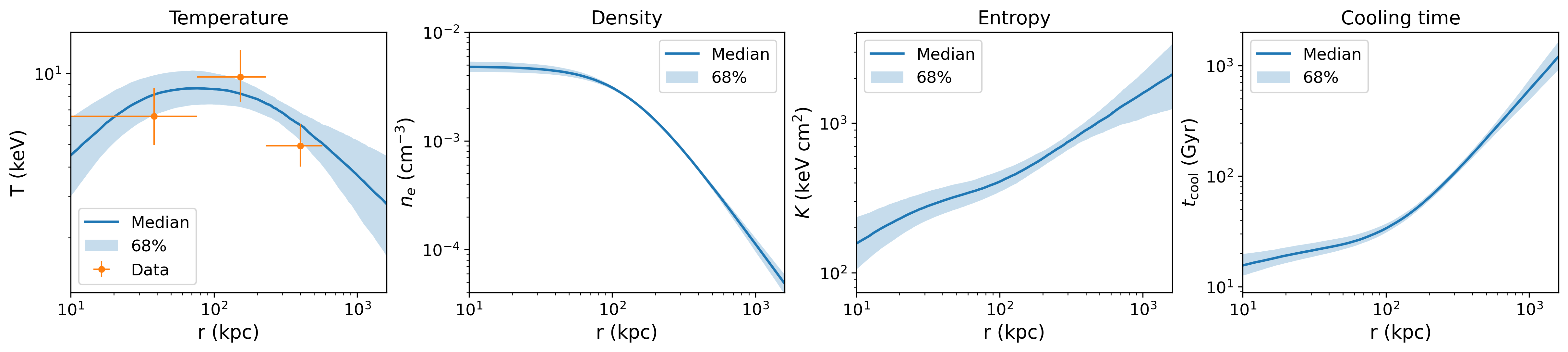}
    \caption{Best-fit temperature, electron number density, entropy, and cooling time profiles with $1\sigma$ confidence intervals derived from the MCMC analysis described in Sect.\ref{sec:cc}.}
    \label{fig:T-ne}
\end{figure*}

\begin{table*}
    \caption{Priors and best-fit parameters for the joint X-ray–SZ MCMC analysis. } 
    \label{tab:mcmc}
    \centering
    \renewcommand{\arraystretch}{1.4}
    \begin{threeparttable}
        \begin{tabular}{lcccccc}
        \hline
        & $n_{0}$  & $r_{\rm c}$ & $\beta$ & $T_{\rm ICM}$ & $ \alpha$ & $S_{\rm bkg}$ \\
        & ($ \rm cm^{-3}$) & (kpc) &  &  (keV)  &   & $(\rm cts~s^{-1}~arcmin^{-2}~cm^{-2})$ \\
        \hline  
        Prior range & $10^{-6} - 0.1$  & $1 - 10^3$   & $0.1 - 2$  & $0.5-20$  & $0-10$  & $10^{-10} - 1$ \\
        Best-fit parameter & $(4.83_{-0.52}^{+0.60}) \times 10^{-3} $ & $125_{-21}^{+25}$ & $ 0.60_{-0.05}^{+0.07} $ & $6.95_{-1.15}^{+1.24}$ & $0.32_{-0.16}^{+0.15}$ & $(1.77\pm 0.02) \times 10^{-5}$ \\  
        \hline
        \end{tabular}
        \tablefoot{Uniform priors are given by their allowed ranges. Best-fit parameters with 1$\sigma$ confidence interval derived from the posterior distributions.}
    \end{threeparttable}
\end{table*}

To investigate the radial temperature structure and assess the presence of a cool core, we extracted spectra from three concentric annuli (see also Figure~\ref{fig:expcorr-img}(b)): an inner region ($<0\arcmin.2$, $\sim 80$~kpc), an intermediate region ($0\arcmin.2$–$0\arcmin.6$), and an outer region ($0\arcmin.6$–$1\arcmin.5$). 
Because of limited photon statistics, the abundances are either less constrained or have large uncertainties. We therefore fixed the abundance to the global value, $Z = 0.61~\rm Z_{\sun}$, derived from the spectra within $0\arcmin -1\arcmin.5$, and refit all three regions. The resulting temperatures changed by less than $7\%$, which is smaller than their statistical uncertainties, indicating that our temperature measurements are robust against the assumed abundance.

Following the approach of \citet{ruppin21}, a joint fit of X-ray and SZ data can better constrain the density profile in low-S/N systems, which allows only one or two radial bins for temperature analysis. The X-ray surface brightness $S_X(R)$ is computed by integrating the gas emissivity along the line of sight:
\begin{equation}
S_{X}(R) = \int_R^{\infty} \Lambda(T,Z) ~ n_{\rm e}(r) ~ n_{\rm p}(r) ~ \frac{r ~dr}{\sqrt{r^2 - R^2}} + S_{\rm bkg},
\end{equation}
where $n_{\rm p}(r)$ ($\approx n_{\rm e}(r) / 1.2$) is the proton density for a fully ionized plasma with one solar abundance, and the cooling function $\Lambda(T,Z)$ was implemented via a lookup table generated with \textsc{xspec} over a temperature grid of $T = 0.5$–20 keV, assuming a constant abundance of $0.61\,Z_\odot$, so that $\Lambda(T,Z)$ can be rapidly evaluated at each temperature.
The integrated Compton-$y$ parameter from SZ measurements probes the electron pressure ($P_{\rm e} = n_{\rm e} ~ T_{\rm e}$):
\begin{equation}
Y_{\rm SZ}^{\theta_{\rm max}} = 2\pi \frac{k_{\rm B} \sigma_{\rm T}}{m_{\rm e} c^2} \int_0^{\theta_{\rm max}} \int n_{\rm e} ~ T_{\rm e} ~ \theta ~ dl ~ d\theta.
\end{equation}
where $\sigma_{\rm T}$ is the Thomson scattering cross section, $m_{\rm e}$ is the mass of the electron, and $c$ is the speed of light. $\theta_{\rm max}$ is the maximum angular distance from the cluster centre considered to integrate the SZ signal. Here, we adopted $Y_{\rm SZ}^{0\arcmin.75}$ $= 5.4\pm 1.2 \times 10^{-5}$~$\rm arcmin^{2}$ from the catalogue in \cite{bleem15}.
Since we have three temperature bins, we included an additional constraint by fitting the temperature profile using the temperature model from \citet{vikhlinin06}:
\begin{equation}
T(x) = 1.35 ~ T_{\rm ICM} ~ \frac{(x/0.045)^{1.9} + \alpha}{(x/0.045)^{1.9} + 1} ~ \frac{1}{\left[1 + (x/0.6)^{2}\right]^{\alpha}},
\end{equation}
where $x = r/R_{500}$, $T_{\rm ICM}$ is the mean ICM temperature, and $\alpha$ is the core-to-mean temperature ratio. The electron density profile $n_{\rm e}(r)$ was modeled with a single-$\beta$ function:
\begin{equation}
n_{\rm e}(r) = n_{0} \left[1 + \left(\frac{r}{r_{\rm c}}\right)^2\right]^{-3\beta/2}.
\end{equation}
where $r_{c}$ and $\beta$ are the core radius and slope parameter.
As described in \cite{ruppin21}, we performed a Markov Chain Monte Carlo (MCMC) analysis using the \texttt{emcee} Python package \citep{foreman13}, using 96 walkers and 15000 steps per walker, to efficiently sample the parameter space of $n_{0}$, $r_c$, $\beta$, $T_{\rm ICM}$, $\alpha$, and $S_{\rm bkg}$. 
We adopted broad, physically motivated uniform priors on all model parameters within the ranges listed in Table~\ref{tab:mcmc}. Gaussian likelihoods were used to compare the models with the observed X-ray surface-brightness profile, temperature profile, and SZ $Y_{\rm SZ}$ value. 
The best-fit parameters are listed in Table~\ref{tab:mcmc}, and the resulting temperature and electron density profiles, with $1\sigma$ confidence intervals, are shown in Figure~\ref{fig:T-ne}.

Since the X-ray surface brightness approaches the background level at $r \gtrsim 1\arcmin.5$, the outer slope of the density profile is poorly constrained. Attempts to fit more complex models, such as a double-$\beta$ or broken-$\beta$ profile, did not improve the fit. Either the second component was degenerate with the first, or the outer parameters were unconstrained. We therefore adopted the single-$\beta$ model, noting that it may not capture the true outer density slope.

To further assess the cool-core status, we derived the entropy profile, $K$ ($= kT n_{\rm e}^{-2/3}$) and cooling time \citep{sarazin88}: 
\begin{equation}
t_{\rm cool} = 8.5 \times 10^{10}~{\rm yr}~ \left(\frac{n_{\rm H}}{10^{-3}~ \rm cm^{-3}}\right)^{-1} ~ \left(\frac{T}{10^8 ~\rm K}\right)^{1/2}
\end{equation}
and presented them in Figure~\ref{fig:T-ne}. The central entropy and cooling time exceed the weak cool-core thresholds ($K_0 < 150~\rm keV~cm^2$ and $t_{\rm cool} < 7.7$~Gyr; \citealt{hudson10,zhang16}), placing this system in the non–cool-core category.
The elevated central entropy, long cooling time, and disturbed X-ray morphology together suggest that SPT-CL~J0217-5014 has likely experienced merger activity, which could have disrupted any pre-existing cool core and prevented the re-establishment of a stable cooling flow.

\section{DESI optical counterparts}
\label{sec:DESI}
%
\begin{figure*}
    \centering
    \includegraphics[width=\textwidth]{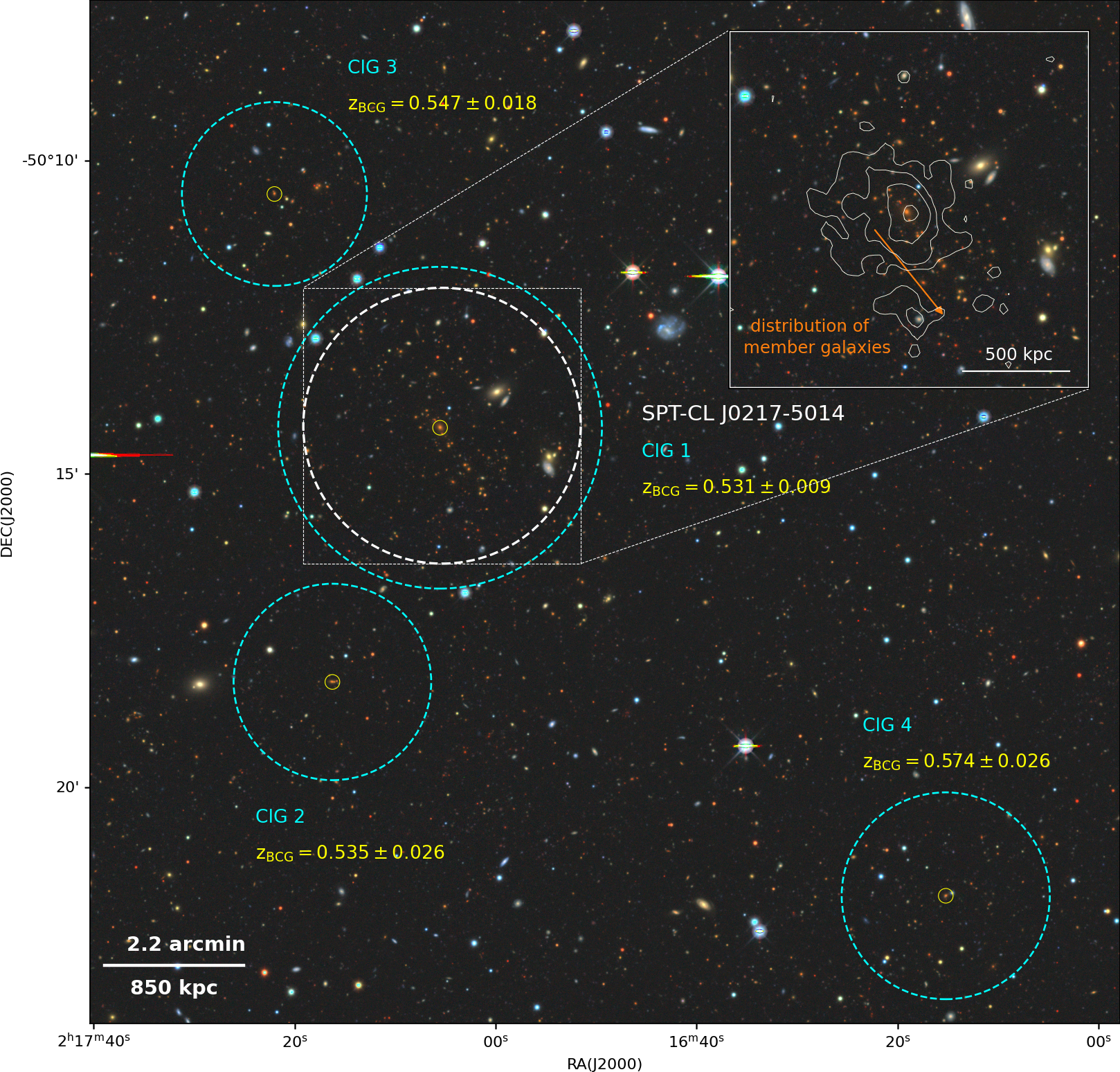}
    \caption{Composite DESI Legacy survey image of SPT-CL J0217-5014 using g-, r-, and z-bands data. The white dashed circle shows the $R_{500}$ region of SPT-CL J0217-5014. Four galaxy clusters identified by \cite{zou22} using DESI DR9 data are marked by cyan dashed circles centred on the brightest cluster galaxies (BCGs) with a radius of $R_{500}$. The BCGs are marked by yellow circles. The photometric redshifts at their BCGs are marked as well. In the inserted zoom-in view of SPT-CL J0217-5014, the X-ray contours shown in Figure~\ref{fig:dyn} are overlaid, and the distribution of member galaxies is indicated.}
    \label{fig:optical}
\end{figure*}

Galaxy clusters form and evolve along the nodes and filaments of the cosmic web, frequently undergoing mergers driven by large-scale structure dynamics. Given the dynamical disturbance indicated by our X-ray analysis (see Section~\ref{sec:dyn}), we examined the optical (and infrared) data for galaxies in the environment of this system to search for potential large-scale structures associated with the cluster.
Several recent large and deep optical surveys, such as the DESI Legacy Imaging Surveys (DR8, DR9, DR10; \citealt{dey19}), DES DR2 \citep{abbott21}, and HSC-SSP PDR3 \citep{aihara22}, have led to the construction of extensive galaxy cluster catalogues \citep[e.g.,][]{zou21,zou22,kluge24,WH24,yantovski24}. Although the spectroscopic survey of DESI has not yet covered the region of our target cluster, photometric data are available. 

We have searched across those publicly available optical cluster catalogues and found that \cite{zou22} identified an optical counterpart of our target cluster, along with three nearby clusters at similar photometric redshifts.  
Using the photometric redshifts of galaxies estimated from DESI DR9 and WISE data, \cite{zou22} applied the Clustering by Fast Search and Find of Density Peaks (CFSFDP) method and identified 532,810 galaxy clusters at $z<1$ with a photo-$z$ uncertainty of about 0.017.  
In Figure~\ref{fig:optical}, we present the DESI Legacy Survey composite image\footnote{\url{https://www.legacysurvey.org/viewer}} of SPT-CL J0217-5014 using g-, r-, and z-bands from DESI DR9. 
Within its $\rm R_{500}$ radius, we find a prominent elongated distribution of red galaxies in the image (highlighted by an orange arrow). This filamentary galaxy distribution is consistent with the X-ray surface brightness excess identified to the south (see Section~\ref{sec:sbp} and Figure~\ref{fig:sbp}).

We notice a small X-ray clump about $1\arcmin.5$ ($\sim 580$~kpc) south of the cluster centre, spatially coincident with a blue galaxy.
According to the photo-z provided by the DESI Legacy Surveys DR9, this blue galaxy is more likely a foreground galaxy at $z = 0.224 \pm 0.025$. To assess its impact, we split the S sector into two equal sectors (see Figure~\ref{fig:expcorr-img}(b)) and extracted the corresponding surface brightness profiles shown in the bottom-right panel of Figure~\ref{fig:sbp}. This foreground galaxy is located approximately $1\arcmin -1\arcmin.5$ from the cluster centre in the S-E sector, and its residual X-ray emission after point-source subtraction contributes part of the excess in this region.
Furthermore, using the S-sector surface brightness profile and the best-fit $\beta$-model, we roughly estimate that the excess emission in the $1\arcmin$–$2\arcmin$ radial range corresponds to a luminosity of $\sim 3.3 \times 10^{42}\ \mathrm{erg\,s^{-1}}$ in the $0.5$–$2$~keV band. This luminosity would be compatible with that of a low-mass galaxy group potentially associated with the blue galaxy; however, a search in the DESI and DES optical catalogues around $z \simeq 0.224$ reveals no galaxy concentration at this position, disfavouring the presence of a foreground galaxy group. Therefore, the foreground galaxy likely accounts for a fraction of the excess in the S–E sector, but not for the entire southern excess.
In contrast, the excess in the S–W sector between $\sim 1\arcmin$ and $2\arcmin$ remains clearly visible and is aligned with the filamentary distribution of red galaxies. This might indicate that the S–W excess is contributed by gas stripped from galaxies falling into SPT-CL J0217–5014 along an intracluster filament connected to the surrounding large-scale cosmic filament, along which external matter is being accreted onto the cluster \citep{maturi13,hyeonghan24}.

Four clusters of galaxies (ClG) identified by \cite{zou22} are marked on the image, with their BCGs' photometric redshifts labelled. 
ClG 1 coincides with our target cluster SPT-CL J0217-5014 and has a richness of 305, while ClG~2–4 have richness values of 61, 49, and 77, respectively. The lower richness of the neighbouring systems indicates that SPT-CL J0217-5014 is the primary, most massive cluster in this complex and likely sits at a node of the surrounding large-scale structure. 
Its disturbed X-ray morphology and the non-cool-core thermodynamic properties point to a dynamical history that has been strongly influenced by mergers. A plausible scenario is that a relatively energetic, possibly close-to-head-on merger in the past could have disrupted a pre-existing cool core and/or prevented the establishment of a long-lived cool core \citep[e.g.,][]{zuhone11,VS21}. In this picture, the western edge would correspond to a merger-driven cold front associated with the remnant low-entropy core of the infalling subcluster, which is now largely embedded within the global X-ray emission of SPT-CL J0217-5014 and therefore not easily identifiable as a separate X-ray peak.

The neighbouring systems, clusters ClG 2 and ClG 3, are located close to SPT-CL J0217-5014 in projection and have similar photometric redshifts. If they are physically associated with the system, they could also represent lower-mass companions that have undergone mergers or infall onto the main cluster a few Gyr ago. 
Such interactions can stir the ICM and excite long-lived sloshing, helping shape the observed cold front and tail-like features \citep[e.g.,][]{hu21}.
At a larger projected distance, another cluster (ClG 4; $z\approx0.57$), lies along the same direction in which the filamentary galaxy distribution within SPT-CL J0217-5014 extends. Although a direct interaction with SPT-CL J0217-5014 is unlikely, its alignment supports the presence of a broader cosmic structure connecting these systems. 
Overall, a combination of past mergers with lower-mass companions and ongoing anisotropic accretion along cosmic filaments provides a natural framework to understand the disturbed ICM and non–cool-core state of SPT-CL~J0217$-$5014.

Although these three nearby clusters lie outside the field of view of \Chandra, the two nearer ones are within the coverage of \Swift observations. However, no significant X-ray counterparts with detectable thermal emission are found in the \Swift data. We also cross-matched with the eROSITA galaxy cluster catalogue \citep{bulbul24}, and find no detections of these systems, likely due to their lower X-ray luminosities. 
Future X-ray observations with higher sensitivity and lower background will be essential to confirm the physical association and thermodynamic properties of these optically identified systems.
The upcoming 4-meter Multi-Object Spectroscopic Telescope (4MOST; \citealt{dejong19}) will conduct a large-field-of-view spectroscopic survey of the southern sky, enabling an independent test of the putative optical filaments linking SPT-CL J0217-5014, as suggested by DESI Legacy Survey imaging. Moreover, the spectroscopic data will reveal the details of the merging process of SPT-CL J0217-5014, providing line-of-sight velocities of member galaxies, mass ratios of subclusters, and an estimate of the merger timescale.

\section{Summary}
\label{sec:summary}

In this work, we present a detailed analysis of the galaxy cluster SPT-CL J0217-5014 using $\sim 100$~ks of \Chandra observations. Our main results are summarised as follows:
\begin{enumerate}
    \item The \Chandra image reveals that SPT-CL J0217-5014 has a disturbed morphology characterised by a surface brightness edge at $\sim 0\arcmin.26$ ($\sim 100$~kpc) to the west and a tail-like feature extending to the east, suggesting a disturbed, non-relaxed ICM.
    \item The cluster's dynamical state was assessed using four morphological parameters. The concentration index ($c$) and centroid shift ($\omega$) are close to the canonical thresholds separating cool-core and non–cool-core systems, and thus do not provide a definitive classification. In contrast, the power ratio ($P_3/P_0$) and morphology index ($\delta$) clearly place SPT-CL J0217-5014 in the dynamically disturbed regime, indicating the system has experienced merger activity.
    \item A joint X-ray and SZ analysis of the surface brightness profile, temperature profile, and the SPT integrated Compton parameter yields the density, entropy, and cooling time profiles. These results classify SPT-CL~J0217-5014 as a non–cool-core cluster.
    \item The best-fit metal abundance within $1\arcmin.5$ ($\sim 0.7\rm R_{500}$) is $0.61_{-0.23}^{+0.26}~\rm Z_{\sun}$. This sub-solar abundance is consistent with the typical metallicities observed in non–cool-core clusters, where dynamical processes could disrupt the cool core and tend to mix the central metal-rich gas with the outer ICM.
    \item Optical data reveal an elongated distribution of red galaxies within the cluster’s $\rm R_{500}$ radius. This filamentary galaxy distribution aligns closely with an X-ray surface brightness excess towards the southern region, possibly due to gas stripped from galaxies infalling along an intracluster filament.
    \item Three potential galaxy clusters ClG 2-4 near SPT-CL J0217-5014 were identified by \cite{zou22} using DESI DR9. Their lower richness confirms that SPT-CL J0217-5014 is the primary, most massive cluster in this complex and likely sits at a node of the surrounding large-scale structure.
    \item SPT-CL J0217–5014 likely underwent a relatively energetic, nearly head-on merger that disrupted a pre-existing cool core; ClG 2 and ClG 3 may be lower-mass companions that have merged with or infallen onto the main cluster, while ClG 4 aligns with the extension of the filamentary galaxy distribution, suggesting its association with a broader cosmic web. Taken together, past mergers and ongoing filamentary accretion provide a natural explanation for the disturbed ICM and the present non–cool-core state.
\end{enumerate}

\begin{acknowledgements}

DH and NW acknowledge the financial support of the GA\v{C}R EXPRO grant No. 21-13491X.  
ZSY is supported by the science research grant from the China Manned Space Project (Grant No. CMS-CSST-2025-A04) and the National SKA Program of China (Grant No. 2022SKA0120103).
SDF, YYZ, and HGX acknowledge the support of the National Natural Science Foundation of China (NFSC) Grant No. 12233005. 

\end{acknowledgements}




%
   \bibliographystyle{aa} 
   \bibliography{reference} 


\begin{appendix} 

\section{Spectral fitting using different background files}
\label{app-1}

\begin{figure}
    \centering
    \includegraphics[scale=.22]{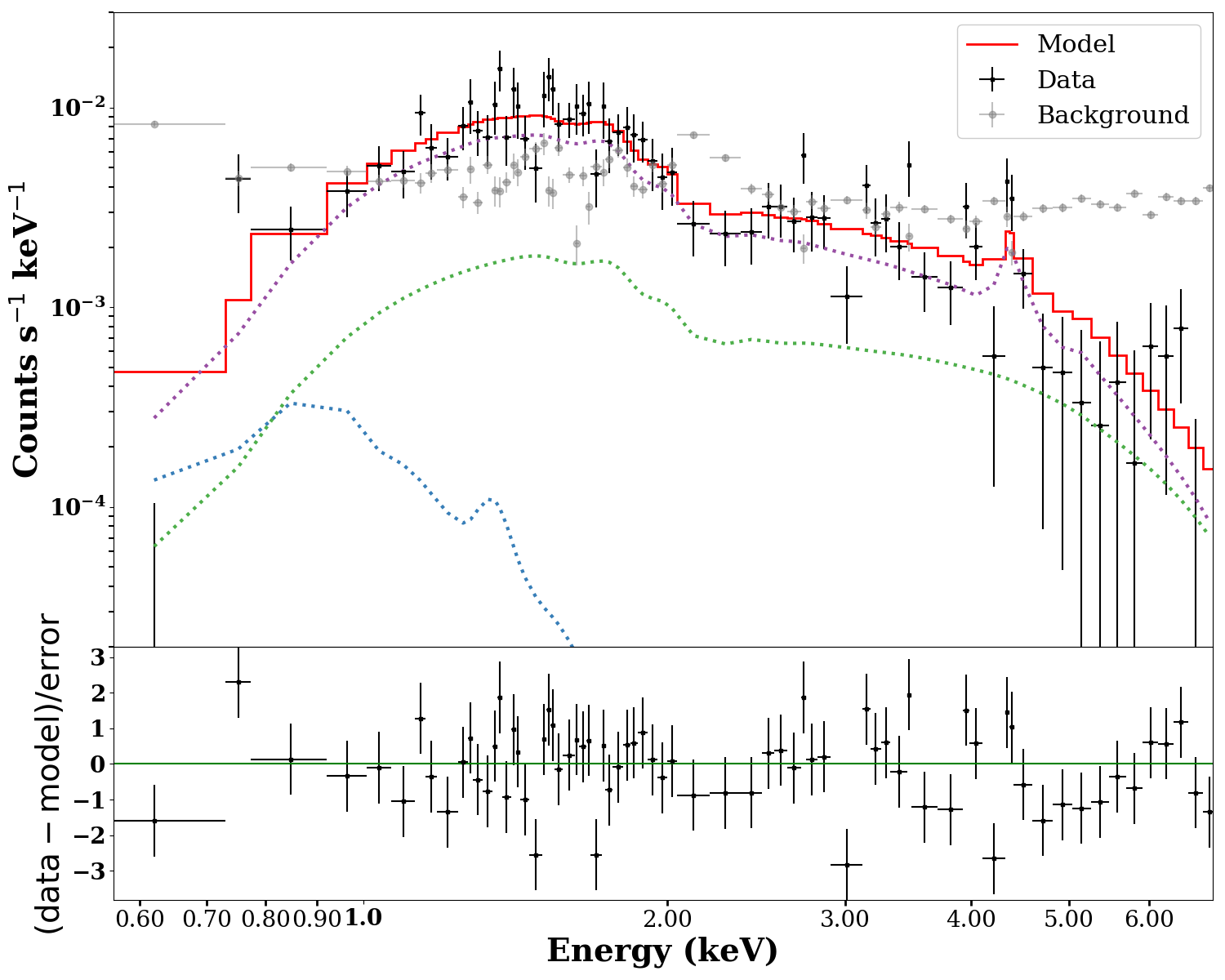}
    \includegraphics[scale=.22]{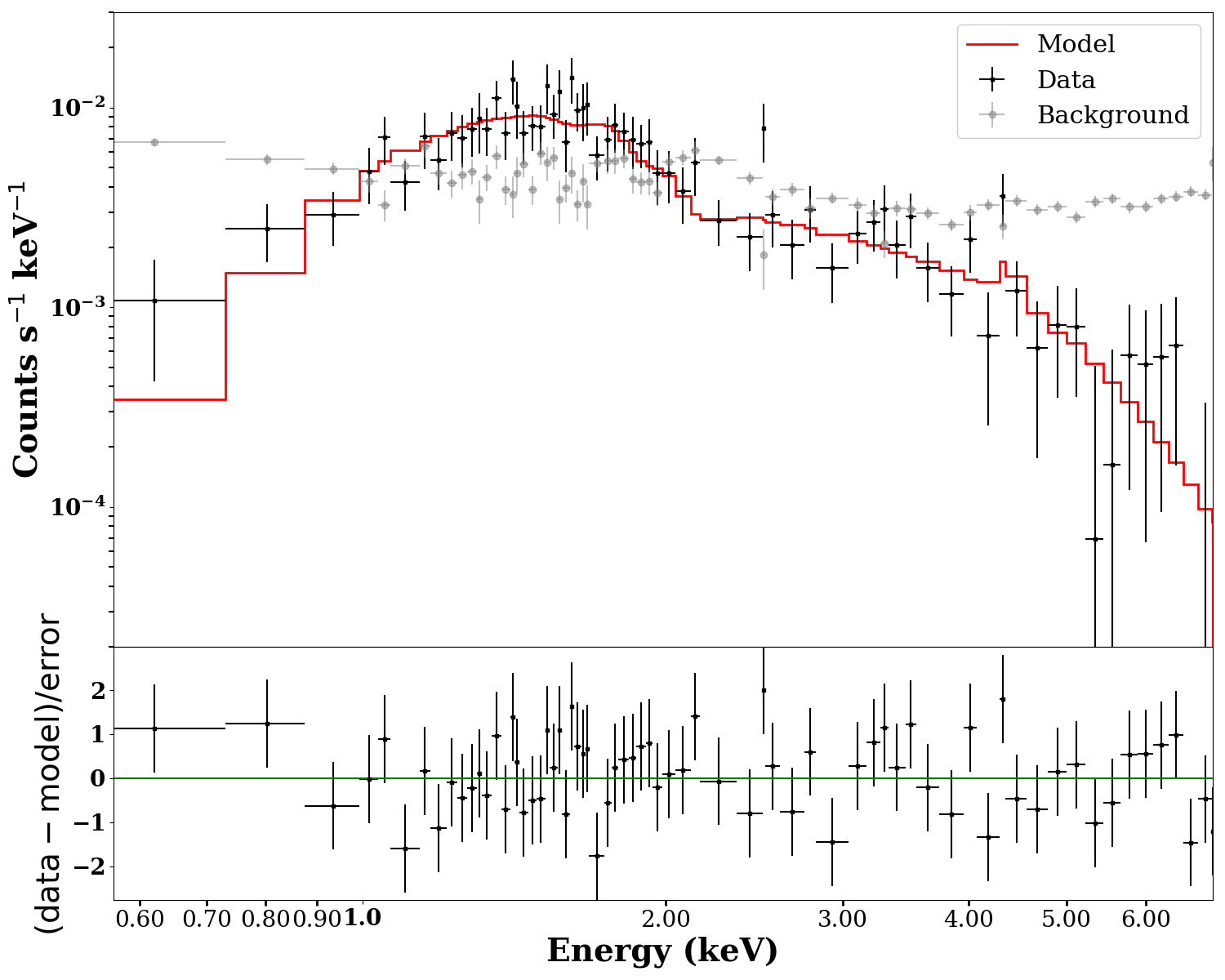}
    \includegraphics[scale=.22]{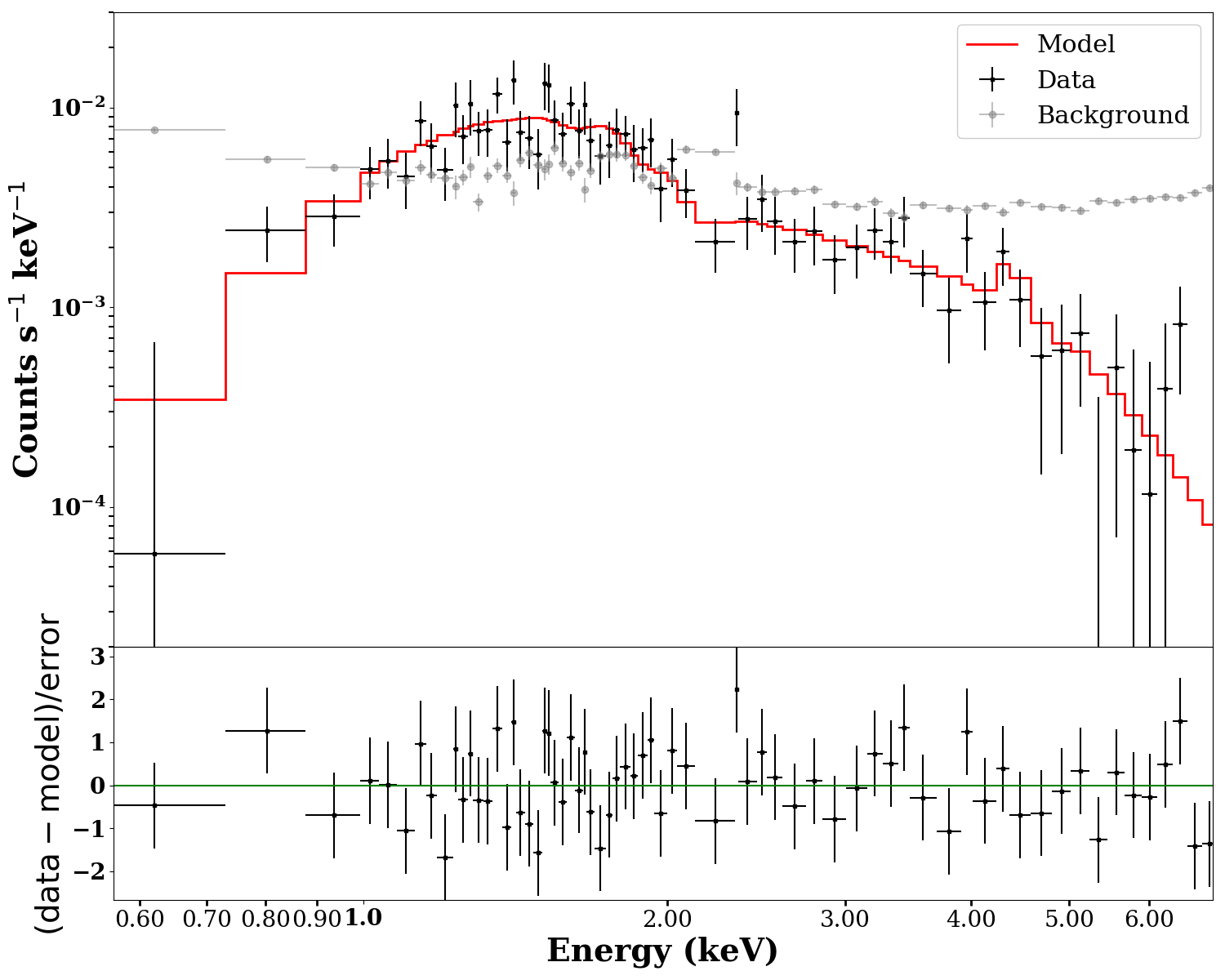}
    \caption{Same as Figure~\ref{fig:spc_fit} but using the other three background files: stowed background (top), local background from annular region (middle), and box regions (bottom).}
    \label{fig:spc_fit_bkg}
\end{figure}

We present the best-fit results using the other three background files, i.e., stowed background and two sets of local backgrounds, in Table~\ref{tab:spc-bkg}, and exhibit their spectra in Figure~\ref{fig:spc_fit_bkg}.
For the spectral fitting using the stowed background, we additionally accounted for the X-ray sky background by adopting the model \texttt{phabs*(apec+powerlaw+apec)}. The first \texttt{apec} component models the soft X-ray emission from the Galactic halo, with its temperature and abundance fixed at 0.2~keV and 1~$\rm Z_{\sun}$ \citep{KS00}, respectively. The \texttt{powerlaw} component represents unresolved cosmic X-ray point sources (e.g., AGNs), with a fixed photon index of $\Gamma = 1.4$ \citep{mushotzky00}.

\begin{table}
    \caption{Spectral fitting results within $1\arcmin.5$ using different background files }
    \label{tab:spc-bkg}
    \centering
    \renewcommand{\arraystretch}{1.3}
    \begin{threeparttable}
        \begin{tabular}{ccccc}
        \hline
        Background  & Temperature & Abundance &  C-stat/dof \\
           & (keV) & ($\rm Z_{\sun}$) & \\
        \hline  
         stowed & $5.76_{-1.89}^{+1.20}$ & $0.50_{-0.27}^{+0.22}  $ & 1139.43/1202 \\
         lbkg\_an & $5.92_{-0.81}^{+1.00}$ &  $0.44_{-0.24}^{+0.26}  $ & 1097.36/1208  \\
         lbkg\_box & $5.54_{-0.78}^{+0.94}$ & $0.36_{-0.24}^{+0.26} $ & 1088.00/1208
         \\
        \hline
        \end{tabular}
    \end{threeparttable}
\end{table}

\end{appendix}

\end{document}